# Photonic Lantern

Sergio G. Leon-Saval*, Alexander Argyros and Joss Bland-Hawthorn

**Abstract:** Multimode optical fibers have been primarily (and almost solely) used as "light pipes" in short distance telecommunications and in remote and astronomical spectroscopy. The modal properties of the multimode waveguides are rarely exploited and mostly discussed in the context of guiding light. Until recently, most photonic applications in the applied sciences have arisen from developments in telecommunications. However, the photonic lantern is one of several devices that arose to solve problems in astrophotonics and space photonics. Interestingly, these devices are now being explored for use in telecommunications and are likely to find commercial use in the next few years, particularly in the development of compact spectrographs. Photonic lanterns allow for a low-loss transformation of a multimode waveguide into a discrete number of single-mode waveguides and vice versa, thus, enabling the use of single-mode photonic technologies in multimode systems. In this review, we will discuss the theory and function of the photonic lantern, along with several different variants of the technology. We will also discuss some of its applications in more detail. Furthermore, we foreshadow future applications of this technology to the field of nanophotonics.

*Corresponding author: Sergio G. Leon-Saval
e-mail: sergio.leon-saval@sydney.edu.au
**Sergio G. Leon-Saval, Alexander Argyros and Joss Bland-Hawthorn:** Institute of Photonics and Optical Science, School of Physics, University of Sydney, Sydney, Australia

# 1 Introduction

Photonics has been described as "molding the flow of light" [1, 2] and, as such, it offers extensive options to manipulate light propagating along a fiber or within a waveguide. Examples include fiber Bragg gratings, array waveguide gratings, optical circulators, fiber couplers, ring resonators, beam shapers, switches, and many more. For the most part, these devices were developed in the context of telecommunications, and their function is restricted to single-mode propagation. Hence multimode systems have not been investigated to the same extent and have been limited in their functionality.

Astrophotonics, a field that lies at the interface of photonics and astronomy, emerged in about 2001 to investigate new enabling technologies for astronomical instrumentation. While optical fibers have been in use in astronomy since the 1980s, they were restricted to merely transporting light between the telescope focus and the instrument. Large-core multimode fibers were used in order to maximize the light captured at the telescope focal plane while matching the field of view of the investigated objects. Subsequently, in the 1990s, single-mode fibers found important uses in astronomical interferometers as a way of filtering the noisy input signal and combining the signal of multiple telescopes or apertures [3-5]. However, for multimode fibers, more complex forms of light manipulation were not considered until only a decade ago in the context of complex filters that were being developed to suppress noise in the form of hundreds of narrow-frequency emission lines from the Earth's atmosphere. The first devices were based on fiber Bragg grating technology and thus could only operate efficiently in single-mode fibers [6]. It was thus recognized that a new photonic device was needed that would transfer properties of single-mode waveguides to a multimode system, interfacing efficiently between single- and multimode fibers. The device that achieved this we now refer to as a photonic lantern and began to emerge soon after [7].

The photonic lantern forms the interface between a multimode waveguide and a set of single-mode waveguides, and thus allows a low-loss transition from one to other as required by the function of the optical system under consideration. Most generally, it consists of a collection of single-mode waveguides (the single-mode (SM) end) that are interfaced to a multimode waveguide (multimode end (MM)) through a physical waveguide transition. In a standard fiber-based photonic lantern, an array of single-mode fibers is placed inside a secondary cladding, of lower index than both the cores and cladding of each of the optical fibers. The transition involves the cores of the SM fibers reducing in size and losing their ability to confine the light. The light thus spreads to the cladding, and becomes confined by the secondary lower index cladding, which have now become core and cladding of the final MM fiber waveguide, respectively.

In this paper we outline the principles that allow photonic lanterns to operate, and discuss the mode transitions that take place, modeling approaches and design considerations. We also present an overview of different implementations of photonic lanterns using different fabrication methods. Although originally developed for Astrophotonics, photonic lanterns are finding uses in other fields that require single- to multimode interfacing, and as such we outline their use in astronomy and space photonics, and telecommunications.

# 2 Theory and modeling

The single-mode end of a photonic lantern comprises an array of isolated identical single-mode waveguides. The array of single-mode waveguides is a multimode system whose spatial modes are the supermodes of the array, which are all degenerate. The number of degenerate supermodes is equal to the number of waveguides [8]. Light can couple between this array of single-mode waveguides and a multimode waveguide via a gradual transition. If the transition is adiabatic, then the supermodes of the single-mode array evolve into the modes of the multimode waveguide, and vice versa. The second law of thermodynamics (brightness theorem) does not allow lossless coupling of light from an arbitrarily excited multimode system into one single-mode system. However if the two systems have the same number of degrees of freedom (i.e. same number of single-mode waveguides as modes in the multimode waveguide), then lossless coupling becomes possible by conserving the entropy of the system. This is a necessary but not sufficient condition. The parameters and properties of ideal low-loss photonic lantern are discussed in more details in the following sections.

## 2.1 Quantum analogy

We consider a lantern with $M$ uncoupled SM waveguides, that transition to a MM fiber with $M$ modes. How the uncoupled SM waveguide modes evolve through an adiabatic transition to become the modes of the MM waveguide can be described by analogy with the Kronig-Penney model [9, 10] for the interaction of electrons in a periodic potential well. At an atomic scale, electrons begin to display wavelike properties, including interference and non-localization. This information is contained within the wavefunction which obeys the wave-equation-like Schrödinger equation. A free electron may be represented by a plane wave, whereas for conduction electrons inside a periodic crystal, as in the Kronig-Penney model, the electron wavefunction is described by Bloch functions.

We can compare the photonic transition that occurs in a photonic lantern between an isolated array of $M$ single-mode waveguides and a multimode one with $M$ modes to a quantum mechanical (QM) system which evolves from $M$ isolated potential wells, each with a single discrete allowed energy level, to a single broader potential with $M$ discrete energy levels. This analogy is depicted in Fig. 1. Figure 1(A) shows the one dimensional refractive index profile ($n$) vs radius widely used to represent optical fibers. However, $1/n$ instead of $n$ is used in the vertical axis as this allows the refractive index profile of the photonic lantern to be represented in the same fashion as the potential ($V$) in the 1D Kronig-Penney model. Thus we compare optical fiber cores with quantum wells as shown in Fig. 1(B).

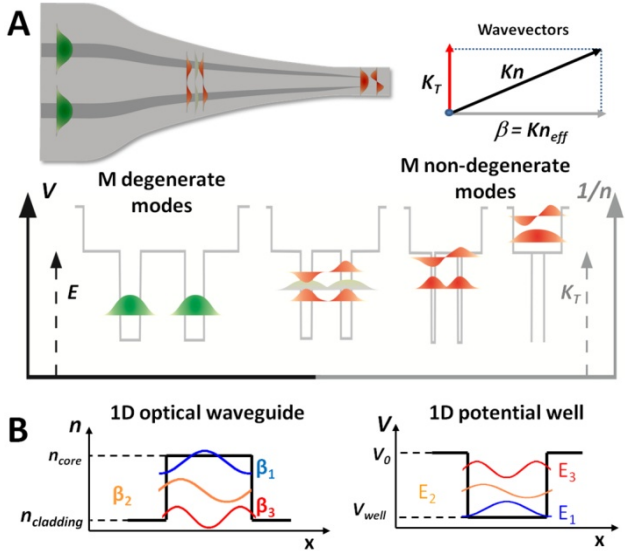

Fig.1. (A) Schematics of the Kronig-Penney model analogy for the photonic lantern. (B) Schematics of a one dimensional quantum well and a one dimensional waveguide.

In the QM case, energy is used to define the wavefunction corresponding to discrete energy levels; however this is not applicable to the electromagnetic (EM) case of spatial modes propagating along a fiber core, as the energy of each photon is fixed ($E = hf$). These spatial modes in the EM case are defined by their propagation constant $\beta = Kn_{eff}$ ($K$ being the wavenumber and $n_{eff}$ the effective index of the mode) and a transverse wavevector $K_T$ (see Fig. 1(A) top right). In this analogy $K_T$ (EM) and $E$ (QM) behave qualitatively the same. We compare the change in energy of the standing wave solutions of the electron inside the quantum wells, with the change in $K_T$ of the spatial modes in the waveguide.

A fiber core can be designed to have only one spatial mode (i.e. the fundamental mode) by choosing the right refractive index ($n$) profile. This mode has its electric field concentrated in the region of high $n$ and hence has the highest modal effective index ($n_{eff}$) and the highest propagation constant $\beta$, hence the lowest transverse wavevector ($K_T$) (see Fig. 1(A)). In the QM case of an isolated potential well, only discrete energies for the electron wavefunction are allowed, and the electron wavefunction takes the form of standing waves. With the right potential and geometry, a potential well can be designed to allow only one discrete energy level (i.e. the ground state). These standing wave solutions of the independent quantum wells have the lowest energy ($E$) and typically their amplitude is concentrated in the regions of low potential ($V$). Hence, the ground state of the potential well becomes analogous to the fundamental mode of the waveguides.

At the start of the transition (Fig.1 (A)), each quantum well allows only one electron in its ground state (fundamental mode). The taper transition renders the quantum wells progressively narrower such that each electron's wavefunction begins to be increasingly less localized, and the energy of each state increases. With the wells closer together, the leaky "conduction" electrons behave as if confined to a periodic crystal. At the point where the taper ends, the wells have essentially vanished, and the collective behaviour of the electrons is described by $M$ states (*cf.* supermodes) confined to a single broad potential well (*cf.* multimode core).

## 2.2 Mode evolution and modeling

In order to understand how a set of identical modes evolve into an equal number of non-degenerate modes, the entire taper transition of the photonic lantern must be modeled. The most effective approach is to model the transition at discrete points along its length, such that at each point the 2D waveguide geometry is considered. The modes at each point may be found by any number of available mode solvers, such as CUDOS MOF Utilities [11], RSoft [12] or COMSOL [13]. The geometry may be constructed and scaled in size, or more practically, the geometry may remain fixed and the wavelength changed by the same scaling factor – increasing the wavelength being equivalent to having a smaller waveguide. This approach makes the implicit assumption that the lantern only varies in scale along the tapered transition, but any deviations from this are unlikely to affect the modes of interest in most cases, and certainly do not affect the mode evolution overall.

For example, air gaps between the single-mode fibers at the single-mode end of the transition (which are not present at the multimode end) will not affect the modes which are highly concentrated in the single-mode cores at that stage.

A second approach is to model the entire three-dimensional transition using methods such as beam propagation (e.g. RSoft). The results of such modeling will depend on the initial conditions used in the model, and thus do not directly give information about the modes and the mode evolution but do allow an exploration of the effect of launch conditions themselves. However, such beam propagation methods may be sensitive to the exact geometry of the structure being modeled, thus the insight derived may be more qualitative than quantitative. Full three-dimensional simulations using other approaches such as finite-element methods may be impractical due to the multi-scale nature of the problem, having a small wavelength and waveguide features compared to the transition length.

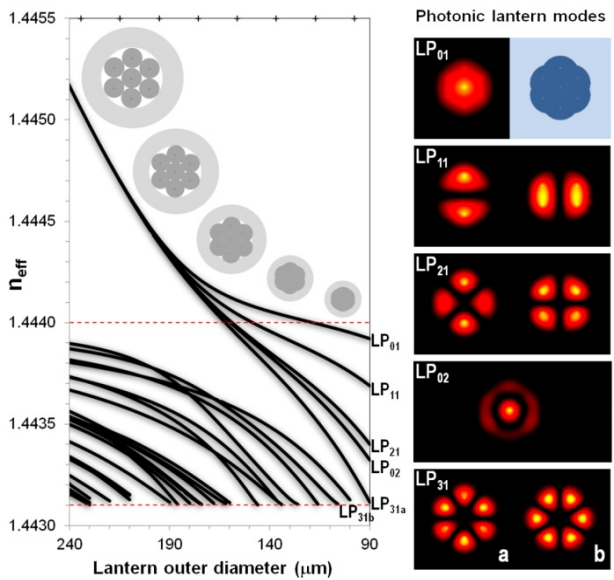

Fig.2. The evolution of modes throughout a 7 fiber photonic lantern. The 7 core modes are degenerate at large diameters, but become non-degenerate at smaller diameters and fill the range of $n_{eff}$ available in the multimode core at the end of the transition. The red horizontal dashed lines indicate the core and cladding index of the final multimode core ($n_{co}$ = 1.444; $n_{cl}$ = 1.4431). (right panel) Detail of the calculated modes supported by the photonic lantern at the multimode end.

Insight into the mode evolution can be gained by the first approach of solving for the modes along the length of the transition. An example of such a study from Ref. [9, 14] is shown in Fig. 2. A 7×1 lantern with a 437 μm diameter at the single-mode end and a 90 μm diameter at the multimode end was modeled using CUDOS MOF Utilities [11] to calculate the effective indices of the modes along the photonic lantern transition. The structure was defined to be equivalent to the multimode end of the transition, with all features in the cross-section approximated as circles as required by the software. The material parameters were set to those corresponding to the chosen wavelength, and the transition was modeled by varying the wavelength rather than the geometry. The mode profile of the final modes of the structure shown in Fig. 2 were calculated using RSoft.

Figure 2 shows the mode evolution for the 7×1 photonic lantern in Ref [14]. At large lantern diameters (not shown) the modes of the single-mode cores are strongly confined, and thus do not couple and remain near-degenerate. In addition, there is a near-continuum of cladding modes. As the lantern diameter decreases, the V-parameter of the single-mode cores also decreases and the modal fields expand, such that interaction between the single-mode cores increases. The resulting coupling leads the formation of non-degenerate supermodes that form from linear superposition of these original 14 modes. In Fig. 2, it can be seen that this coupling becomes significant at diameters below 190 μm. These modes continue to separate in effective index and eventually become the modes of the multimode core at the end of the transition. At the same time, the previous near-continuum of cladding modes is eliminated as all the modes reach cut-off; in Fig. 2 this occurs at an outer diameter of 100 μm.

The number of single-mode cores determines the number of modes of interest, which must be conserved across the transition in a functioning lantern. In the example of Fig. 2, the 7 single-mode cores and the multimode core both supports 14 modes (including degeneracies). The symmetry of the structure also plays a role and in this case turns the otherwise 4-fold degenerate $LP_{31}$ mode into two 2-fold degenerate pairs $LP_{31a}$ and $LP_{31b}$, one aligned with the cores, and the other with the regions between them. The former remains, but the latter is cutoff to give the desired number of modes in the multimode end of the transition. One final remark is that the fundamental mode (LP$_{01}$) is ordinarily expected to asymptote at the cladding index (1.4440), as it has no cut-off. In this case, the presence of the external cladding results in its effective index reducing below this value.

Apart from revealing the evolution of the modes, Fig. 2 also contains information about the wavelength response of the lantern. Since it depicts the modes along the transition at a single wavelength, the modes at another wavelength would behave identically, once the appropriate scaling was accounted for. For example, a larger wavelength would exhibit the same evolution at correspondingly larger lantern diameters. This neglects only material dispersion.

The mode transition seen in Fig. 2 is in some sense ideal, in that (i) all the cladding modes have been cut-off by the end of the transition, meaning the multimode end of the lantern has the correct number of modes to match the single-mode end, and (ii) there are no crossings between the core modes coming from the single-mode end and the cladding modes, which would create opportunities for loss through power transfer between the desired and undesired modes. However, the above conditions are necessary, rather than sufficient and losses can still occur. The total length of the transition must also be taken into account when designing a lantern. In every device in which cross-sectional geometry changes along the propagation length, the change in the mode field of a propagating mode must be gradual for low-loss transmission. As such, it has been demonstrated numerically by the Beam Propagation Method that the losses of a photonic lantern depend on the tapering angle along the transition length [15, 16]. The optimal transition length, i.e. tapering angle, will also depend on the number of modes in the photonic lantern. As in the case of conventional optical fiber tapers, the adiabaticity criterion is more severe for higher order modes [17, 18]. This can be seen quantitatively in the published results in which a 3-mode lantern was shown to required a tapering angle of < 0.69° [16]; whilst a 85-mode lantern required a tapering angle of < 0.21° [15] to achieve low transmission losses.

## 2.3 Optimum waveguide geometry

So far in published results, photonic lantern core geometries have consisted of hexagonal lattices, rounded array or square lattices with the number of cores approximately equal to the number of final waveguide modes as required [14, 15, 19-21]. However, it has been theoretically demonstrated that the best starting point for a low loss lantern design is an uncoupled core geometry that best approximates or samples the geometry of the modes of the final MM fiber [16, 22, 23]. Hence, the ideal arrangements and number of single-mode cores which minimize losses are unique for each photonic lantern and depend on the number of modes required in each case.

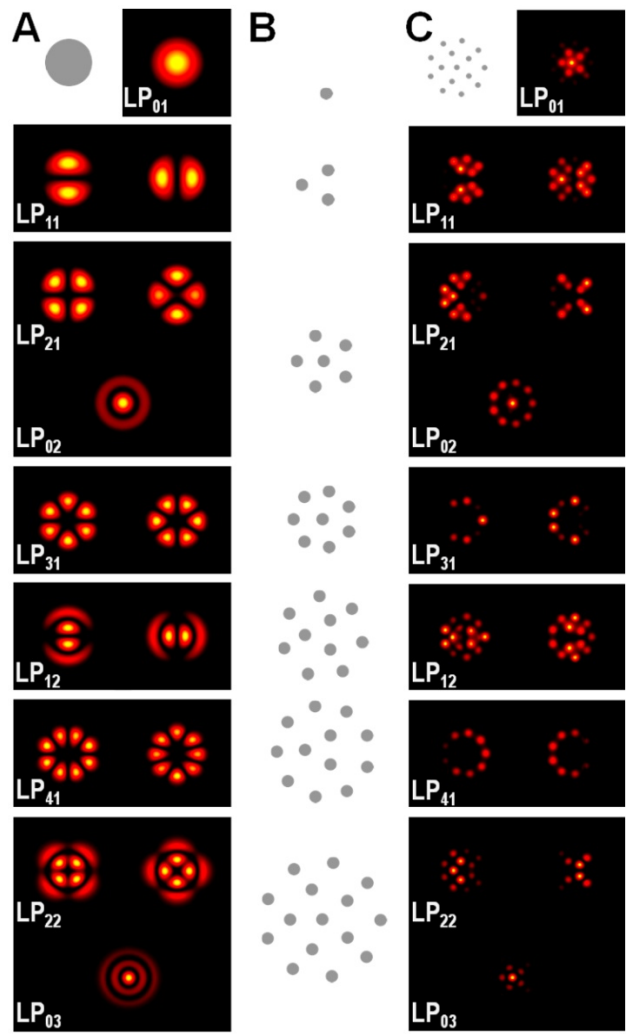

Fig.3. (A) 15 lowest order step index fiber spatial modes. (B) Coupled waveguide arrays whose supermodes closely match the fiber modes. (C) 15 lowest order modes of the near optimal 15 core arrangement.

When the cores are isolated or weakly coupled, their supermodes (or superposition of supermodes) should resemble the final multimode waveguide modes. In a step-index MMF that supports the LP$_{lm}$ mode, $l$ is the maximum azimuthal node number, and $m$ is the radial node number. To best approximate this mode shape, the geometry of the uncoupled single-mode waveguides must have $m$ concentric rings, with each ring having an odd number of waveguides so as to differentiate between the cos($l\phi$) and sin($l\phi$) azimuthal dependence of the (degenerate) LP$_{lm}$ modes. The number of waveguides in each ring is required to be $2p+1$ where $p$ is the largest $l$ for each radial mode number, $m$.

For example, Fig. 3(A) shows the 15 lowest order LP modes for a standard step-index MMF, and each panel groups the modes with almost identical cut-off frequency. Fig. 3(B) shows the waveguide patterns that best approximate a MMF supporting 3, 6, 8, 10, 12, and 15 spatial modes in total, and Fig. 3(C) shows the corresponding supermodes of the 15 core array.

The exact coupling from the modes in the single-mode end to the those of the multimode end and vice versa is in practice extremely complicated, and will depend on the amplitude and phase of the light in each mode or waveguide, the transition, and any perturbations or imperfections in the transition. In an adiabatic photonic lantern with no perturbations, light launched in a particular supermode as in Fig. 3 (C); i.e. launched into the corresponding set of single-mode waveguides with the correct intensity and phase, will end up in the matching MM fiber mode. Under normal operating conditions, light launched into the multimode end will couple into a set of MM fiber guided modes and then into a set of different SM waveguides with different intensities (through the corresponding superposition of supermodes). Thus standard photonic lanterns are not "true" mode (de)multiplexers and do not couple each mode of the MM waveguide into just one of the SM waveguides. However, with a careful waveguide design and geometry a photonic lantern functioning as a "true" mode (de)multiplexer (i.e. mode sorter) into the different SM waveguides could be achieved.

## 2.4 Wavelength dependence

The properties and efficiencies of these multimode to single-mode converters are indeed wavelength dependant. A essential condition for a low loss device is that the number of modes in the multimode section $M_m$ and the number of modes in the single-mode section $M_{sm}$ must be the same: $M_m=M_{sm}$. $M_{sm}$ is fixed by the number of single-mode cores, and does not change with wavelength so long as the cores remain single-moded, however $M_m$ is wavelength dependent. The approximate number of unpolarized spatial modes guided by a MMF with a step-index profile is given by

$$M_m \approx \frac{Vp_m^2}{4} = \frac{\pi^2 d_m^2 NA_m^2}{4\lambda^2} \quad (1)$$

where $Vp_m$ is the V-parameter from waveguide theory [14]; $d_m$ is the MM fiber core diameter; and $NA_m$ is the numerical aperture of the waveguide for a given wavelength $\lambda$.

A mismatch in the number of modes ($M_m \neq M_{sm}$) will increase the transmission losses of the lantern. Light could couple to modes that become cutoff in the SM to MM direction when $M_m < M_{sm}$ and similarly, when $M_m > M_{sm}$ in the MM to SM direction. Thus, a perfect lantern device has an intrinsically limited bandwidth, since the number of modes could only be perfectly matched over a restricted wavelength ranges determined by the modal cut-off of the multimode waveguide at those wavelengths and the number of single-mode waveguides. Devices with a 300 nm bandwidth have been already demonstrated with a minimum 85% transmission at the edges of the usable wavelength span [15, 24].

## 3 Different approaches

The only way that these mode converters can be made is by making a physical transition in which the single-mode waveguides either stop acting as such, and/or cease behaving as independent uncoupled waveguides. The final aim of this physical transition is to adiabatically form a multimode waveguide in which the single-mode waveguides either vanish or form a composite waveguide formed by strong cou-pling between them. Photonic lanterns to date have been manufactured and demonstrated by three different techniques; two using optical fibers [14, 15] and the third by using ultrafast laser writing techniques [20, 25].

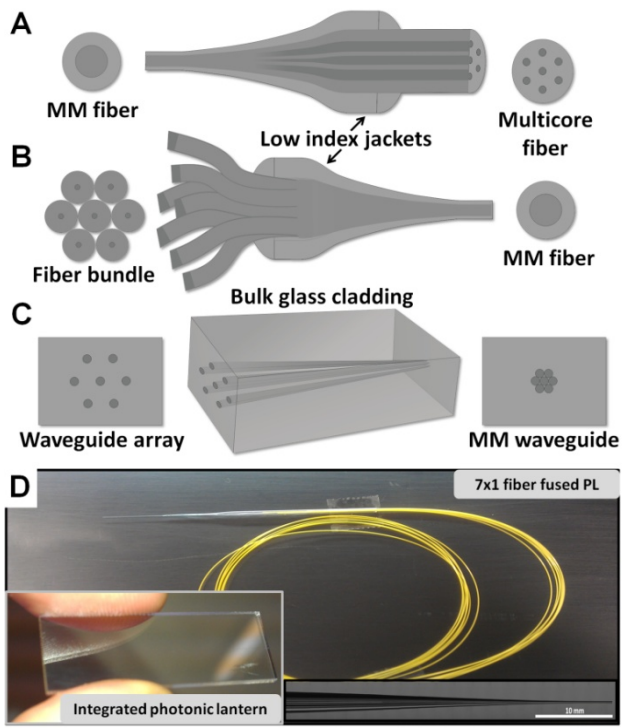

Fig.4. Schematics of the three different photonic lantern fabrication approaches. (A) Multicore fiber; (B) Fiber bundle; and (C) Ultrafast laser inscription (integrated photonic lantern). (D) Optical photograph of a 7×1 fiber fused photonic lantern and (bottom left panel) an integrated photonic lantern fabricated by Dr R. Thomson (Ref [20]).

The very first photonic lantern, reported in 2005 [7], was fabricated by using a drawing tower fabrication method for photonic crystal fibers devices. This method, called the 'ferrule technique' [26], was the precursor of the photonic lanterns and was used to interface different photonic crystal fibers to single-mode standard step-index fibers. Nowadays, an all solid optical fiber splitter/combiner fabrication technique [14] is most commonly used. A bundle of single-mode fibers is inserted into a low index glass capillary tube which is then fused and tapered down in a glass processing machine to form an all solid multimode fiber at the other end (Fig. 4 (B) and (D)). The second approach using optical fibers for the realization of photonic lanterns is by using a multicore fiber with an array of identical single-mode cores [15]. A photonic lantern can be made by tapering such a multicore fiber whilst, again, placing a low refractive index jacket around the fiber to form the cladding of the multimode fiber (Fig. 4 A). Each method has its advantages and disadvantages. The multicore fiber approach has the potential to produce photonic lanterns with a much larger number of modes in the system in a more straight forward manner. A multicore fiber with hundreds of cores can be fabricated in the same fashion as an all-solid photonic bandgap fiber [27], for example. Furthermore, the fabrication procedure for a multicore photonic lantern – one single fiber inside the low index jacket – is simpler and less prone to fabrication imperfections during the tapering process. However, when input/output are considered and/or a different photonic function is required for each single-mode core, then multicore photonic lanterns are not the appropriate choice. Photonic lanterns produced by the fiber fused bundle approach with large number of fibers are cumbersome to fabricate. Nevertheless, this approach produces lanterns that are easily interfaced and connectorized to any standard single-mode fiber system, opening a broad range of possible applications.

An entirely different approach for the fabrication of photonic lanterns is based on ultrafast laser inscription (ULI) techniques. Here, laser writing is used to create waveguides in a piece of bulk glass (Fig. 4 C). The laser illumination causes an increase in the refractive index, which forms the waveguide core, and the technique allows for positioning of these cores in 3D within the bulk glass [28, 29]. In this case, the isolated single-mode waveguide cores are created and gradually brought together such as they couple strongly. This creates the adiabatic optical transition required for the single-mode to multimode conversion, and the strongly coupled cores form the final multimode composite waveguide [20, 21, 25]. ULI photonic lanterns, also called integrated photonic lanterns, are a very versatile approach. Almost any single-mode waveguide geometry can be produced at the same time as offering the possibility to scale to very large numbers of waveguides. Their compact footprint and versatility makes them a very attractive approach for the implementation of photonic lanterns, however, interfacing to standard single-mode systems and the transmission losses need to be addressed.

# 4 Applications

## 4.1 Astronomy and Spectroscopy

Multimode fibers are widely used in optical and infrared astronomy to allow many celestial sources to be observed simultaneously. These large-core fibers (50-300 μm core diameter) are used to transport light from the telescope focal plane to a remote spectrograph. Mainstream astronomy has generally avoided single-mode fibers because of the difficulty of coupling light into these efficiently. Even the best performing adaptive optics systems, which attempt to deliver a diffraction-limited beam, are unable to couple light efficiently into the Gaussian-like mode of single-mode fibers below 2500 nm [3, 30, 31]. Consequently, astronomers have been unable to exploit numerous technological advances in photonics over the past three decades, as these have almost entirely been based on single-mode telecommunications fiber.

There are two major drivers in astronomy for the photonic lantern technology: one is the use of single-mode photonic technologies such as Fiber Bragg Gratings (FBG) [32] and Arrayed Waveguide Gratings [33, 34], and the second is a push towards more compact, stable and precise next generation astronomical spectrographs working in the diffraction limited regime [35, 36] and hence requiring single-mode fiber inputs. The latter also has direct applications to small satellite instrumentation and exoplanets research.

An early successful application of photonic lanterns has been to suppress unwanted noise sources in astronomical observations. An OH-sky suppression instrument called GNOSIS [24], installed at the 3.9 meters Anglo Australian Telescope (AAT) in Australia (Fig. 5 (A)), was made possible by photonic lantern technology. This instrument efficiently transfers the light captured and coupled into multimode fibers by the telescope into single-mode fibers through a photonic lantern. These are spliced to FBGs which suppress the OH emission lines emitted by the night sky, followed by another photonic lantern transforming the filtered light back into a multimode fiber before entering the telescope spectrograph. This improves the signal to noise ratio, having removed the noise on the way to the spectrograph and thus the quality of the scientific observations. It also highlights the fundamental premise behind the photonic lantern itself, in efficiently interfacing a multimode requirement (for light capturing efficiency) and a single-mode requirement (for the FBG light filtering).

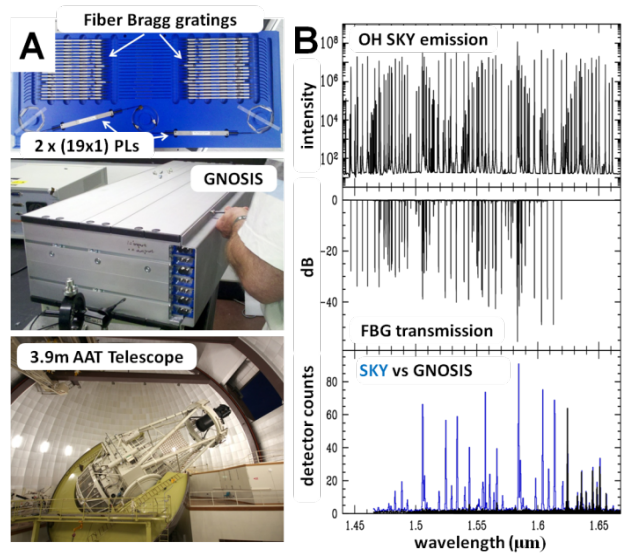

Fig.5. (A) Optical photographs of the built GNOSIS instrument: (top) FBG filter tray with two 19×1 photonic lanterns spliced back to back with the OH FBG filters between them; (middle) Fully assembled GNOSIS FBG unit which consists of 7 trays like the one in the top panel image; and (bottom) the 3.9 m Anglo-Australian Telescope in Australia where the GNOSIS instrument was installed and tested. (B) On sky results of the GNOSIS instrument. Bottom plot shows a comparison of the filtered and unfiltered OH sky background.

In order to exploit the full capabilities of this OH suppression concept new exciting approaches are being pursued recently with a follow up astronomical instrument called PRAXIS. Multicore fibers with 37, 55 and 120 cores are being produced for the fabrication of OH sky filters. There, the standard fiber fused photonic lanterns will be replaced with multicore photonic lanterns. The advantage will be in the ability to inscribe the gratings in all the single-mode cores (the multicore fiber) simultaneously, rather than in each single-mode fiber separately. However, the major technological challenge in this concept is the realization of high quality fiber Bragg gratings in multicore fibers with such large numbers of cores, but recent pub-lished results on the fabrication of fiber Bragg gratings in multicore fibers [15, 37] have been promising.

A further application relates to the spectrographs themselves. One of the fundamental design characteristics of a spectrograph is the amount of light collected, dispersed and re-imaged onto a detector. Spectrographs operating at high spectral resolution ($R=\lambda/\Delta\lambda > 20,000$) are the most challenging. In low light applications there is a strong tension between the need to broaden the spectrograph input (i.e. a larger slit entrance, to allow for more light) and preserving the spectroscopic performance of the instrument (which requires a narrower slit). In 2010, the Photonic Integrated Multi-Mode Spectrograph (PIMMS) was proposed; this promised a fundamental shift in fiber-fed spectrograph design [35, 36]. In this approach, the photonic lantern conversion from MM to SM is used to completely decouple the spectrograph design and performance from the light source at the MM input. Photonic lanterns can achieve throughputs equivalent to a MM fiber design with a spectral resolution of a diffraction-limited slit width, which is provided by the SM fibers. The spectrograph is designed to match the output of the array of SM fibers (whose output remains fundamentally unchanged regardless of the source at the MM input), reducing its complexity and size. This reduction in size and optical components has opened a new path towards extremely high resolution, high accuracy, and portable spectrographs. Two areas where this approach is being proposed and studied currently are in high accuracy spectrographs for exoplanet research [38] and space satellite spectroscopy [39].

Extremely precise Doppler spectroscopy is an indispensable tool to find and characterize earth-like planets; however, nearly one order of magnitude better radial velocity (RV) precision over the best current spectrographs [40, 41] is required. The two major limiting factors for achieving this extremely high precision is the multimode nature of the fiber feeding the spectrograph, and the stability of the system. Transforming the multimode input into a single-mode diffraction limited slit by means of a photonic lantern eliminates and reduces, respectively, both of those problems. Firstly, by eliminating the MM fiber input at the spectrographs slit the Point Spread Function instability due to fiber illumination, guiding or scrambling will be removed [42]. Secondly, the diffraction limited input to the spectrograph from the SM fiber allows for a very compact instrument design (as shown in Fig. 6 B) providing a much easier system to control and deliver high optomechanical stability.

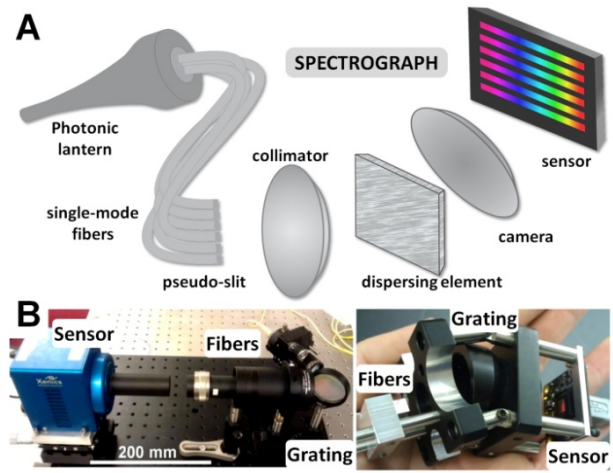

Fig.6. (A) Schematics of a single-mode fiber fed diffraction-limited spectrograph constructed with a photonic lantern. (B) (left panel) IR spectrograph with a optical resolution of R~30,000 (~0.045nm) at 1550 nm. (right panel) Extremely hand-palm compact size diffraction limited VIS spectrograph with an optical resolution of R~1,100 (~0.44 nm) at 550 nm.

The applications in space photonics arise from limitations in payload technology of small satellites due to their restricted mass and sizes. Earth observation and scientific missions are still a relatively minor application of small satellite technology, and the use of such satellites is primarily dominated by telecommunications. However, in recent years there has been a definite trend towards putting small things together to achieve big accomplishments. Photonic lanterns have been proposed for space-borne spectroscopy instrumentation as a solution to the mass and size limitations. By using photonic lanterns, medium to high-resolution spectrographs can be designed to fit in smaller spaces and with less optical components compared to their standard multimode fiber-fed counterparts. Small prototypes spectrographs have already been developed and tested in balloon launches. Fig. 6 (B) left panel, shows a < 90 g diffraction limited spectrograph fed by 8 single-mode fibers with a resolution of 0.45 nm and an operating wavelength range of 400 to 700 nm.

Photonic lanterns are also been recently studied in astronomy for their modal mixing and scrambling properties [15, 43]. These multimode/single-mode hybrid systems show interesting modal scrambling properties for use in exoplanets research and astronomical instrumentation were low NA fiber delivery is required. Furthermore, photonic lanterns can be used as NA convertors, i.e. focal ratio convertors; asymmetric photonic lanterns can be fabricated such as both ends of a multimode/single-mode/multimode device could in principle have different core sizes and input/output NAs, while conserving the etendue of the system.

## 4.2 Telecommunications

The data capacity carried by a single-mode fiber is rapidly approaching its limits [44, 45]. Multiplexing has been used to increase the capacity of a single SM fiber, with various approaches including wavelength division multiplexing (WDM) and polarization. Spatial Division Multiplexing (SDM) [46] is another approach investigated that has recently being advanced by the research community, albeit being proposed many years ago. This approach can further address this capacity crunch by using a MM (few moded) fiber instead of a SM fiber, and using its spatial modes [47-50] and even high order orbital angular momentum modes in specialty fiber [51] as an additional degree of freedom to increase the number of data channels. Another SDM approach is the use of multicore fibers [52-54], here the spatial multiplexing is done by using the uncoupled cores of a multicore fiber as independent data channels.

Recently, photonic lanterns have been studied as adiabatic mode converters (i.e. as mode-multiplexers and mode-demultiplexers) for their use in SDM systems for coherent multiple-input multiple-output (MIMO) networks [55, 56]. A MIMO system consists of a spatial multiplexer to couple $T$ channels to $M$ fiber modes, a MM fiber with $M$ supported modes, an $M$ mode-demultiplexer to couple to an array of $R$ coherent receivers (Coh. Rx), and electrical MIMO processing (Fig. 7 (A)). Maximal capacity in systems with $T = M = R$ can occur only when the mode-dependent losses (MDL) are negligible [55]. Some spatial multiplexers directly excite the spatial modes and others excite an orthogonal combination of modes. In free beam systems due to the use of passive beam-combiners, phase masks, and spatial-light modulators the coupling losses (CPL) increase proportionally with the number of modes in the system. Thus, spatial multiplexers supporting large number of modes with negligible MDL and CPL are highly desirable. Photonic lanterns can be used as both the spatial multiplexer and spatial demultiplexer in this context. Optimizing the geometrical arrangement of the single-mode waveguides, as discussed in Section 2.3 along the taper transition could reduce the MDL and CPL to zero. Launching into the isolated single-mode core modes distributes the information across the MM fiber modes, this could mitigate the effect of MDL along the transmission. This approach will excite an orthogonal combination of modes such that all channels experience similar modal dependencies reducing the outage probability [57-59].

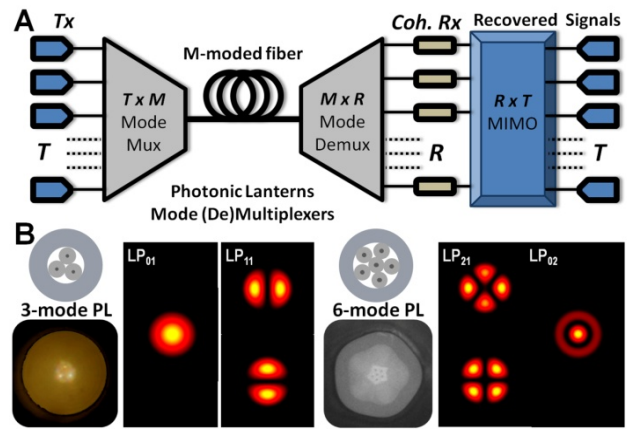

Fig.7. (A) Coherent SDM with electronic MIMO processing ($T<=M<=R$). (B) 3- and 6-moded photonic lanterns mode multiplexers for low-loss SDM coupling fabricated at the University of Sydney, optical photographs and calculated modes.

A further approach currently under development is the use of photonic lanterns to produce mode sorters, i.e. true mode multiplexers. It is challenging to build spatial multiplexers for few-mode fiber that excites each mode without loss since the M spatial modes of the few-mode fiber are spatially overlapping and cannot be simply separated. For instance, phase-mask spatial multiplexers convert Gaussian beams into the different spatial modes, then overlap the shaped beams onto the few-mode fiber using passive beam combining [58]. These multiplexers can have in excess of 20-dB mode selectivity, but suffer from large insertion losses from the passive splitting that increases proportionally to M. In the contrary, photonic lanterns can be nearly lossless, can scale to many modes, and they can be spliced directly o the few-mode fiber as well as easily integrated through the SM fiber pigtails. Standard photonic lanterns have no mode-selectivity since they scramble each signal launched on the SM fiber end onto a combination of the few-mode fiber modes. However, it is possible to add mode-selectivity to the photonic lantern spatial multiplexer by building the lantern with M dissimilar SM fibers rather than M identical fibers. During the adiabatic taper, the dissimilarity can control the coupling between the cores and force certain cores to evolve into specific mode groups (e.g., $LP_{01}$ or $LP_{11}$) [60].

## 4.3 Multimode Plasmonics

A very fast-moving area on Nanophotonics is the study of surface plasmons (SPs) [61], a field known as Plasmonics. SPs are extremely sensitive to any change in the material parameters and/or surface topology. In addition, the exponential decay is usually so strong that most SPs exhibit transverse, sub-wavelength confinement. Due to these properties, SPs have been extensively studied in recent years and are considered especially attractive for enhancing optical phenomena, detecting biochemical events [62] and controlling light at the nanoscale [61]. The field of Plasmonics offers the potential towards higher integration densities for optical circuits by combining the high capacity in photonics and the miniaturization technologies in electronics.

Photonic lantern technology and concepts could be exploited to produce for example beam splitters which are an important functionality for integrated optical circuits. Several concepts in plasmonics have already been developed, e.g., Y-shaped splitters [63], and splitters related to the multimode interference (MMI) [64], however analog photonic functionalities are still far from developed. For instance, the concept of multimode SPs existence and excitation is still an open area of research. It has been shown recently that a metal surface strip can support SP polariton modes, leaky modes with phase constants which are close to those of a SP polariton travelling along an extended thin film. This study demonstrated that the propagation of light along surface plasmon waveguides is mediated by a discrete number of guided polariton modes as well as a continuum of radiation modes [65]. At a given frequency and for a sufficiently wide SP waveguide (with several SP polariton eigenmodes), one could achieve a multimode SP polariton excitation. In fact a strong indication for this are the multimode interference studies [64], which provide direct evidence for multiple guided modes. A SP photonic lantern device could indeed help to study this exciting new direction for plasmonics and nanophotonics. Furthermore, this multimode to single-mode convertors could be used for clean excitation and/or splitting of SP waveguides enhancing even further the SP capabilities for controlling light at the nanoscale.

# 5 Conclusions

The photonic lantern is a versatile and powerful concept, allowing the transformation of an optical multimode system into a single-mode one and enabling the use of single-mode-based photonic technologies in multimode systems for the first time. Photonic lanterns increase the functionality and possible applications of few-moded devices and systems. We have presented the operating principle of these devices and how they can be fabricated with current and standard waveguide technologies. Furthermore, the key current applications for this multimode to single-mode convertors have been presented. Being a relatively recent development, photonic lanterns are likely to find uses in other emerging fields. These mode convertors offer the possibility of improving the light collecting ability while keeping and opening new photonic functionalities. Applications such as gas or Raman spectroscopy as well as coherent detection sys-tems like for example light detection and ranging (LIDAR), are areas in which photonic lanterns could in principle revolutionize existing instrumentation and applications.

# Acknowledgements

The authors acknowledge the contribution of many people to the development and implementation of this technology over its relatively short life span. A special acknowledgement has to go to Tim Birks, another one of the pioneers of this technology; Danny Noordegraaf, for his pioneering demonstration of standard fiber fused lanterns; and to Nick Fontaine, for opening the door to the telecom applications.